%% file: main.tex
\documentclass[conference, 11pt]{IEEEtran}

\usepackage{placeins}
\usepackage[noadjust]{cite}
\input{include}

\input{custom_commands}

\IEEEoverridecommandlockouts

\begin{document}
\title{Differentiable Point Scattering Models for Efficient Radar Target Characterization}
\author{\IEEEauthorblockN{Zachary Chance, Adam Kern, Arianna Burch, Justin Goodwin}\IEEEauthorblockA{MIT Lincoln Laboratory\\Lexington, MA 02421\\Email: zachary.chance@ll.mit.edu, adam.kern@ll.mit.edu, arianna.burch@ll.mit.edu, jgoodwin@ll.mit.edu}}
\date{}

\maketitle

\begin{abstract}

Target characterization is an important step in many defense missions, often relying on fitting a known target model to observed data. Optimization of model parameters can be computationally expensive depending on the model complexity, thus having models that both describe the data well and that can be efficiently optimized is critical. This work introduces a class of radar models that can be used to represent the radar scattering response of a target at high frequencies while also enabling the use of gradient-based optimization.

\end{abstract}

\section{Introduction}

Many defense applications involve the estimation of a quantity of interest, e.g., target position and velocity, or learning patterns in data, e.g., target classification from observed features. Algorithms used to solve these problems often perform optimization in order to fit a chosen model to the available data based on a goodness-of-fit metric. For instance, this can be fitting a ballistic trajectory to a series of detections to maximize a likelihood function \cite{Chang80} or fitting a neural network to correctly label a target embedded in radar data to minimize misclassification error~\cite{blasch2020review}. Depending on the complexity of the model, however, this optimization may require substantial time, computation, or other resources. Thus, efficiently fitting models to data is an important and common task within many defense missions.

A popular class of methods for expedient model fitting is gradient-based optimization, which uses local derivative information from the goodness-of-fit to intelligently refine the model parameters. More specifically, the gradient is a measure of the best direction to perturb the model parameters in order to increase the goodness-of-fit. This is key to the efficiency of gradient-based methods as utilizing the gradient can alleviate expensive searches for better feasible solutions. The primary necessity for this class of optimization techniques is the ability to calculate the derivatives of the goodness-of-fit relative to the model parameters. In many practical applications, it can be difficult to find models that satisfy this requirement; thus, this work will provide a differentiable (i.e., able to provide a gradient) model to enable gradient-based optimization in an important defense mission area: radar target characterization.

Radar target modeling techniques depend on two major factors: the relative size of the target \(l\) to the radar wavelength \(\lambda\) and whether to use parametric or nonparametric models~\cite{Uluisik08}.
Modeling in the Rayleigh region, when \(l \ll \lambda\), is simplified as the entire target contributes to the RCS as a single point scatterer. For modeling in the resonance regime, where \(l \approx \lambda\), nonparametric numeric methods are often used. In this case, the target's geometry and materials all contribute to its RCS as a whole. Popular numeric methods include the Method of Moments~\cite{Ishimaru17}, the Fast Multipole Method~\cite{Harrington68}, the Finite-Difference-Time-Domain Method~\cite{Grel03}, and the Transmission-Line-Matrix Method~\cite{Kunz93}. These techniques provide exact solutions to Maxwell's equations, creating very accurate models of target scattering.

In defense applications, however, targets are often illuminated using high-frequency radar waveforms. This leads to the optical regime, where \(l \gg \lambda\). In this case, scattering can often be reduced to a summation from discrete scattering centers by taking advantage of the Geometric Theory of Diffraction~\cite{Keller62}. This allows the use of point scattering models that model targets parametrically~\cite{Ross68, Carriere92, Potter95, Jackson08} and reduces the problem of radar target modeling to effectively reproducing or simulating the signature of each scatterer. A number of methods have been developed to determine the values of these parameters from radar observations. For instance, Singular Value Decomposition and total least squares were used to approximate both the number of scatterers and their parameters from RCS measurements of the target~\cite{Carriere92}. In \cite{Potter95}, a technique is proposed that uses an iterative approach that cycles between optimizing different parameters for each scatterer via the Nelder-Mead (downhill simplex) method. Similarly, \cite{vanDerMerwe97} also uses the Nelder-Mead method to iteratively optimize model parameters after breaking the data into smaller cells.

Another major contrast among radar signature models is the validity region over viewing angles. For instance, most of the early parametric modelling work focused on a single perspective and extracting scattering center information in range only \cite{Ross68, Carriere92, Potter95}. Models, mostly for the purposes of synthetic aperture radar, expanded upon this to allow for target components that can vary both in range and aspect angle \cite{vanDerMerwe97, Jackson08, Cebula04}. These models can be powerful as they allow for the combination of multiple viewing geometries when fitting to data. For this work, focus will be on models that are valid over all aspect angles.

Notably, none of the aforementioned methods take advantage of gradient information to fit the parameters of the model. To this end, a point scattering model defined over all aspect angles is proposed that can be used to describe the predominant scattering centers of a target over all viewing angles. Further, the proposed model is designed to allow for optimization techniques utilizing gradient-based methods; therefore, enabling efficient fitting to collected radar measurements. Efficacy of the model and its use in estimation and learning applications will be demonstrated via simulated radar data. Contributions of this work are:
\begin{itemize}
    \item Description of differentiable point scattering models and derivation of their general gradient form for range profiles
    \item Demonstration of utilizing differentiable point scattering models for model estimation using gradient descent
    \item Analysis of the advantages and disadvantages of coherent and noncoherent loss functions for model parameter estimation
\end{itemize}

The structure of the paper is as follows: Section~\ref{sec:scattering} describes a point scattering model for radar targets and Section~\ref{sec:range-profiles} describes target range profiles. Sections~\ref{sec:estimation} and \ref{sec:gradients} derive the equations for the gradients of coherent and noncoherent loss functions with respect to point scattering model parameters. Specifically, Section~\ref{sec:estimation} derives the gradients of loss functions with respect to range profiles and Section~\ref{sec:gradients} derives the gradients of range profiles with respect to model parameters, both of which are needed for gradient descent. Section~\ref{sec:estimation} also provides equations for calculating the Cram\'er-Rao lower bound for the point scattering model. Finally, in Section~\ref{sec:results} several experiments are run to show the feasibility of model parameter estimation using gradient descent for both single range profiles and collections of range profiles and compare the effectiveness of coherent and noncoherent loss functions for model parameter estimation.

\section{Point Scattering Models}\label{sec:scattering}
For a rigid body whose perspective is slowly changing (with respect to the duration of a radar pulse), its scattering response can be closely modeled as a linear time-invariant system. Thus, the reflected electric field can be described as the convolution of the radar waveform, \(x(t)\), with a target impulse response, \(h(t)\) \cite{Oppenheim97, Levanon88}. More specifically, the received signal, \(y(t, \bell)\), is given by
\begin{equation*}
y(t, \bell) = x(t)e^{j2\pi f_c t} \ast h(t, \bell),
\end{equation*}
\noindent where \(f_c\) is the center frequency of transmission and \(\bell\) is a line-of-sight vector describing the viewing perspective to the target. A scattering model is a mapping from a real-valued parameter vector \(\btheta\) to a synthesized target impulse response, \(h(t, \bell; \btheta)\).

\begin{figure}
  \centering
  \includegraphics[width=0.85\columnwidth]{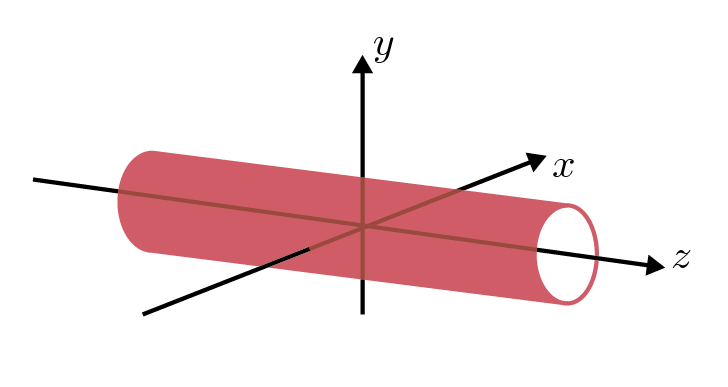}
  \caption{Target coordinate system.}\label{fig:target_coord_xyz}
\end{figure}

\begin{figure}
  \centering
  \includegraphics[width=0.95\columnwidth]{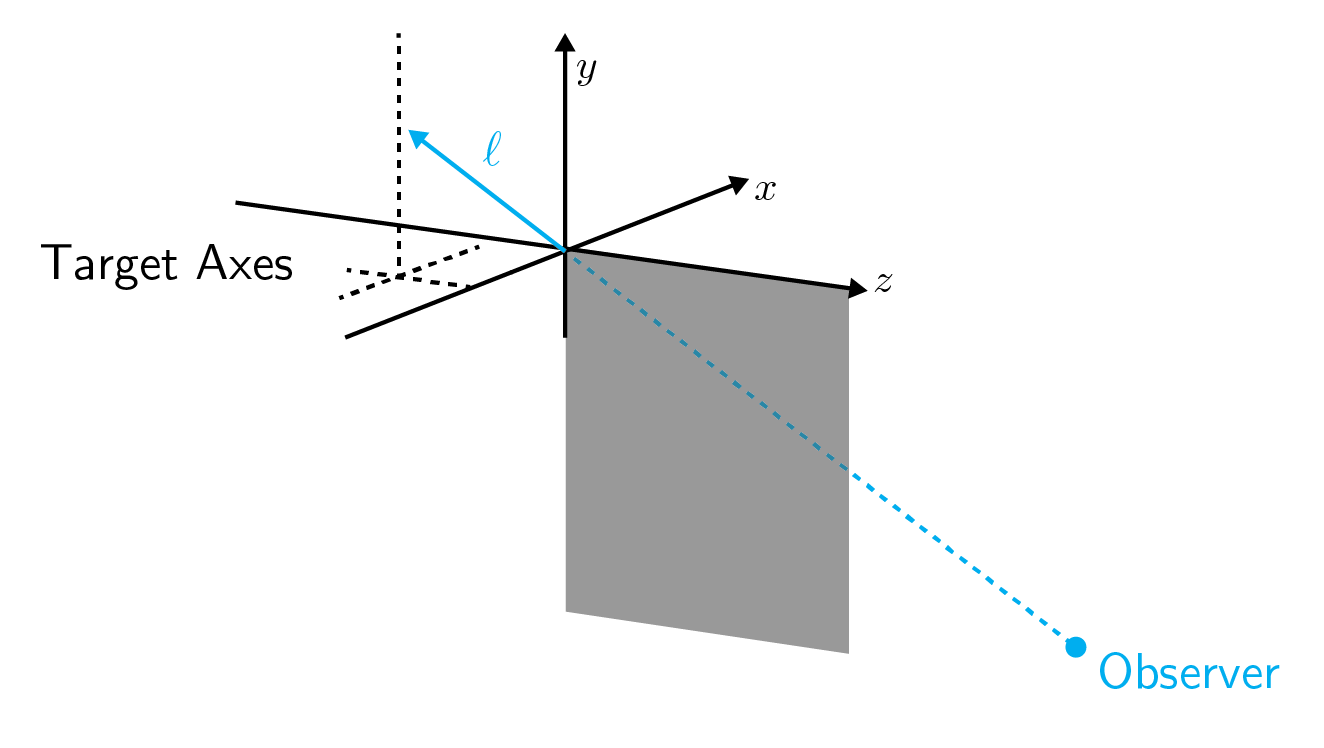}
  \caption{Target line-of-sight vector.}\label{fig:target_coord_los}
\end{figure}

To formally introduce the line-of-sight vector, \(\bell\), consider a Cartesian coordinate system attached to the target as in Figure \ref{fig:target_coord_xyz}. The location of the observer in this system will be denoted as \(\bp_s\). Likewise, there is an observed target at a point \(\bp_t\). It is then possible to define a line-of-sight vector, \(\bell = [\ell_x, \ell_y, \ell_z]^T\), of unit length that points away from the observer toward the target (see Figure \ref{fig:target_coord_los}). Mathematically, the line-of-sight vector is given by
\[
\bell = \frac{\bp_t - \bp_s}{\|\bp_t - \bp_s\|}.
\]

For this work, the class of composite scattering models will be considered. A composite scattering model is created by summing constituent scattering model components such that the total impulse response is given by
\begin{equation}
h(t, \bell; \btheta) = \displaystyle \sum_{n = 1}^N h_n(t, \bell; \btheta_n),
\end{equation}
\noindent where $h_n(t, \bell; \btheta_n)$ is the impulse response for the $n^{\textrm{th}}$ scattering model component, $\btheta_n$ is the parameter vector for the $n^{\textrm{th}}$ scattering model component, and the total parameter vector is $\btheta = [\btheta_1, \btheta_2, \ldots, \btheta_N]^T$.

A composite scattering model that is commonly used due to its simplicity is the \emph{point scattering model}. A point scattering model represents a target as an assemblage of scatterers, each with infinitesimal extent. Each scatterer in the model is completely described using a scattering amplitude\footnote{To account for polarization of the impinging waveform, the amplitude of a scatterer is defined by a four element scattering matrix. For the sake of simplicity, this work will focus on one of the four scattering coefficients; i.e., $a_n$ is a single entry in a scattering matrix.} and location. Let the amplitude of the $n^{\textrm{th}}$ scatterer be denoted as $a_n$. Likewise, its location (a vector in a Cartesian frame attached to the target) will be $\bp_n$. In general, the amplitudes, $a_n$, and locations, $\bp_n$, can be functions of viewing geometry and the parameter vector, $\btheta_n$. For a point scattering model, each component impulse response is given as a Dirac delta function with a defined delay and amplitude such that
\begin{equation}
h_n(t, \bell; \btheta_n) = a_n\delta(t - d_n),\label{eq:pointimp}
\end{equation}
\noindent where $\delta(t)$ is the Dirac delta function, and $d_n$ is the propagation delay to the $n^{\textrm{th}}$ point scatterer. Note that this formulation assumes that the amplitude of each point scatterer exhibits little variation over the observed bandwidth; this assumption may not be fitting depending on the expected shape features. The propagation delay can be expanded as
\[
d_n = \frac{2r_n}{c},
\]
\noindent where $r_n$ is the range to the $n^{\textrm{th}}$ scatterer (from a defined reference point) and $c$ is the speed of light. The range can be computed (assuming far field geometry) as
\[
r_n = -\bp^T_n\bell.
\]
In short, the models considered in this work will be of the following form:
\begin{equation}
h(t, \bell; \btheta) = \displaystyle \sum_{n = 1}^N a_n\delta(t + 2\bp_n^T\bell/c).\label{eq:ptmdl_imp}
\end{equation}
\noindent Note that the frequency response of the model at line-of-sight $\bell$ can be obtained by taking the Fourier transform of (\ref{eq:ptmdl_imp}) to get
\[
H(f, \bell; \btheta) = \displaystyle \sum_{n = 1}^N a_n e^{j4\pi f \bp_n^T\bell/c}.
\]

To form the main focus of the paper, the scope will be narrowed to a particularly useful subset of point scattering models whose amplitudes and positions are differentiable with respect to model parameters; this class of model is defined below.

\begin{definition}{Differentiable Point Scattering Model}.
A differentiable point scattering model is a point scattering model described by the following impulse response:
\begin{equation}
h(t, \bell; \btheta) = \displaystyle \sum_{n = 1}^N a_n\delta(t + 2\bp_n^T\bell/c),
\end{equation}
\noindent where $a_n$ and $\bp_n$ are partially differentiable with respect to the model parameter vector $\btheta$ for all $n = 1, 2, \ldots, N$, i.e., $\partial a_n/\partial \btheta$ and $\partial \bp_n/\partial \btheta$ exist.
\end{definition}

\noindent The class of differentiable point scattering models offer many advantages for optimization and estimation performance characterization. These benefits will be shown throughout the remainder of the paper.

\section{Target Range Profiles}\label{sec:range-profiles}

Before describing scattering model estimation and optimization, it is necessary to describe the form of observed data. A common view of a target from a radar perspective is the shape of the matched filtered pulse in the target vicinity, commonly called a range profile. The primary focus of this work will be fitting a point scattering model to a set of range profile vectors. A range profile of a target is constructed by: (i) removing the carrier frequency component $e^{j2\pi f_c t}$, (ii) matched filtering the received signal, $y(t, \bell)$, and (iii) mapping from time to range via $r = ct/2$ where $c$ is the speed of light. To begin, matched filtering is performed by convolving the received signal with a time-reversed version of the transmit waveform, $x(t)$ \cite{KayDet93}. Thus, the received signal after matched filtering, $g(t, \bell)$, is given by
\begin{align}
g(t, \bell) &= y(t, \bell)e^{-j2\pi f_c t} \ast x(-t),\nonumber\\
&= (x(t)e^{j2\pi f_c t} \ast h(t, \bell))e^{-j2\pi f_c t} \ast x(-t),\nonumber\\
&= x(t) \ast h(t, \bell)e^{-j2\pi f_c t} \ast x(-t),\nonumber\\
&= x(t) \ast x(-t) \ast h(t, \bell)e^{-j2\pi f_c t},\nonumber\\
&= R_{xx}(t) \ast h(t, \bell)e^{-j2\pi f_c t}, \label{eq:matched}
\end{align}
\noindent where $R_{xx}(\tau)$ is the autocorrelation function of the transmit waveform such that
\[
R_{xx}(\tau) = \displaystyle\int_{-\infty}^{\infty}x(u)x^*(u - \tau)du.
\]
\noindent Finally, the range profile, $g(r, \bell)$ is obtained by substituting $t = 2r/c$ into (\ref{eq:matched}), producing
\begin{equation}
g(r, \bell) = R_{xx}(2r/c) \ast h(2r/c, \bell)e^{-j4\pi f_c r/c}. \label{eq:rangeprof}
\end{equation}
\noindent The general process of radar target estimation and learning utilizing range profile vectors will now be discussed, so that the role of the target scattering model (and its derivative properties) will become evident.

\begin{figure}
  \centering
  \includegraphics[width=0.95\columnwidth]{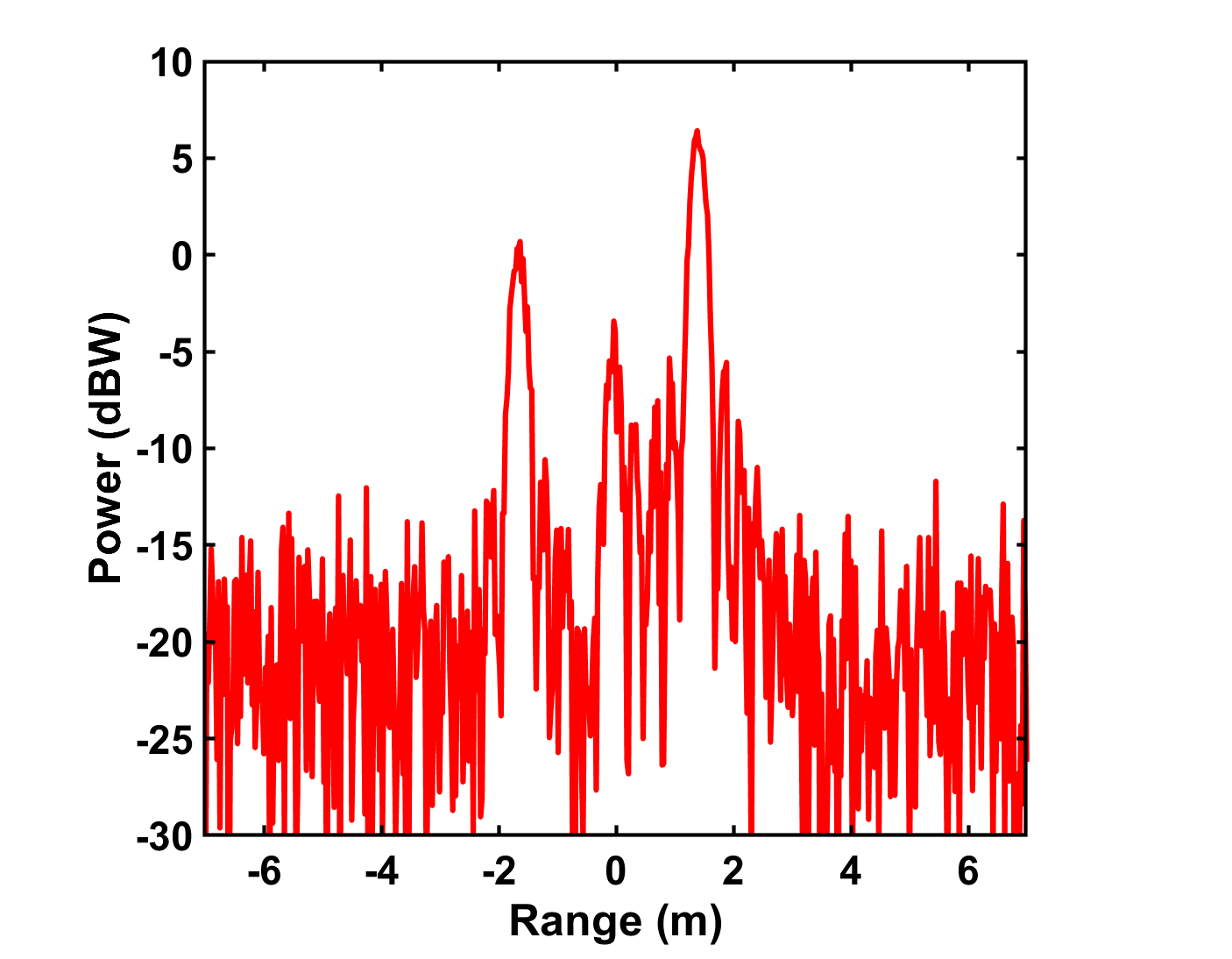}
  \caption{Example range profile, $\bg(\bell)$, of a target.}\label{fig:ex_prof}
\end{figure}

In practice, one typically does not have access to a continuous version of the range profile, $g(r, \bell)$, but a sampled version instead. To introduce this, let the number of samples be $M$ and the range bin size be denoted $\Delta$, then the sampled range profile can be compactly described as a column vector
\[
\bg(\bell) = \left[g(b_1, \bell), g(b_2, \bell), \ldots, g(b_M, \bell)\right]^T,
\]
\noindent for range bins $b_k = (k - 1)\Delta + b_0$ and initial range bin $b_0$. A plot of an example range profile is shown in Figure \ref{fig:ex_prof}.

The range profile of a point scattering model can be obtained by substituting the impulse response (\ref{eq:ptmdl_imp}) into (\ref{eq:rangeprof}) and getting
\begin{equation}
g(r, \bell) = \displaystyle\sum_{n = 1}^N \gamma_n a_n R_{xx}\left(2r/c + 2\bp_n^T\bell/c\right), \label{eq:rangeprof2}
\end{equation}
\noindent where
\begin{equation}
\gamma_n = e^{j4\pi f_c \bp_n^T\bell/c}\label{eq:gamma}
\end{equation}
is the phase delay to the $n^{\textrm{th}}$ point scatterer, and the dependence of $\gamma_n$, $a_n$, and $d_n$ on $\bell$ and $\btheta_n$ is implicit.

Finally, observations of the target will inherently be noisy; thus, each measurement vector, \(\bz\), will be given as
\begin{equation}
\bz = \bg(\bell) + \bw,
\end{equation}
\noindent where $\bw$ is complex Gaussian noise distributed as $\bw \sim \cC\cN(\bzeros, \bR)$. With the form of the data having been defined, performing estimation of a model's parameters from noisy observations will be discussed.

\section{Target Model Estimation}\label{sec:estimation}

Many estimation algorithms require the optimization of a loss function over model parameters, where loss functions are typically chosen to minimize an expected error of the estimator, e.g., minimum mean square error \cite{KayEst93}. More specifically, it is possible to write the target model estimation problem as
\begin{equation}
\hat{\btheta} = \argmin_{\btheta} L(\bz, \bg(\bell; \btheta))\label{eq:loss}
\end{equation}
\noindent for a chosen real-valued loss function \(L(\bz, \bg)\) with noisy range profile vector, \(\bz\), hypothesized range profile vector, \(\bg(\bell; \btheta)\), and model parameter vector, \(\btheta\). Note that this formulation also allows the same optimization to extend past individual range profiles to collections of range profiles, as discussed in Section~\ref{sec:results}.

A common class of algorithms for solving the optimization problem in (\ref{eq:loss}) is gradient-based methods, which uses local derivative information to iteratively refine a solution. For instance, gradient-based methods are also often used for general maximum-likelihood techniques such as iterative least squares. The main caveat for the use of gradient-based methods is that they require the ability to take partial derivatives of the loss function with respect to the model parameters.

One complication introduced by working with radar data is the fact that they are typically complex-valued; this requires more care be taken when deriving the partial derivatives of the loss with respect to the model parameters \cite{KreutzDelgado09}. More specifically, in problems involving complex data, the loss function can be explicitly written as a function of the real and imaginary parts of the data and hypothesized data vectors, $L(\bz_r, \bz_i, \bg_r, \bg_i)$, such that $\bz = \bz_r + j\bz_i$ and $\bg = \bg_r + j\bg_i$. Then, the gradient of the loss function, with respect to the model parameters, can be derived as
\begin{align}
\nabla_\theta L &= \bG_r^T\nabla_{g_r}L + \bG_i^T\nabla_{g_i}L,\label{eq:loss_grad1}
\end{align}
\noindent where $\bG_r$ is the real part of the Jacobian of the hypothesized data vector, $\bg_r$, with respect to the parameter vector, $\btheta$, such that
\[
\bG_r = \Re\left\{\frac{\partial \bg(\btheta)}{\partial\btheta}\right\},
\]
\noindent and similarly
\[
\bG_i = \Im\left\{\frac{\partial \bg(\btheta)}{\partial\btheta}\right\}.
\]
\noindent Note that the gradient of loss (\ref{eq:loss_grad1}) has two sets of contributing terms:
\begin{itemize}
    \item Jacobian of the hypothesized data vector with respect to the model parameters, $\bG$; this is dictated by the form of the data and the chosen model
    \item Gradients of the loss function with respect to real and imaginary parts of the hypothesized data vector, $\nabla_{g_r}L$ and $\nabla_{g_i}L$; this is defined by the choice of loss function
\end{itemize}
\noindent It is assumed that the gradients of the loss function are known to exist, as choice of loss function is outside of the scope of this work. To demonstrate the use of differentiable point scattering models in estimation, two common loss functions will now be discussed along with their respective gradients with respect to the hypothesized range profile, \(\bg(\bell; \btheta)\).

\subsection{Coherent Weighted Squared Error Loss}
Due to its common use, the specific case of a weighted squared error loss is studied. In more detail, the loss function can be written as
\begin{align}
L(\bz, \bg) &= (\bz - \bg)^H \bW (\bz - \bg),\nonumber\\
&= \bz^H\bW\bz - 2\Re\left\{\bz^H\bW\bg\right\} + \bg^H\bW\bg,\nonumber\\
&= (\bz_r - j\bz_i)^T\bW(\bz_r + j\bz_i)~\nonumber - \\
&\phantom{= }~2\Re\left\{(\bz_r - j\bz_i)^T\bW(\bg_r + j\bg_i)\right\}~+\nonumber\\
&\phantom{= }~(\bg_r - j\bg_i)^T\bW(\bg_r + j\bg_i).\label{eq:wls1}
\end{align}
\noindent Note that this form assumes that the weight matrix, $\bW$, is real-valued and symmetric. Now, using (\ref{eq:wls1}), the gradient of the loss with respect to the real and imaginary parts of the hypothesized data vector, $\bg_r$ and $\bg_i$, are
\begin{align}
\nabla_{g_r}L &= 2\bW(\bg_r - \bz_r),\nonumber\\
\nabla_{g_i}L &= 2\bW(\bg_i - \bz_i).\nonumber
\end{align}
\noindent Therefore, employing (\ref{eq:loss_grad1}) the gradient of the loss function with respect to the model parameter vector, $\btheta$, is
\[
\nabla_\theta L = 2\bG_r^T\bW(\bg_r - \bz_r) + 2\bG_i^T\bW(\bg_i - \bz_i).
\]

\subsection{Noncoherent Weighted Squared Error Loss}
If phase information is unreliable, it may be preferable to operate on the amplitudes of the data and hypothesized model. Consider the vector function $\bu(\bx)$ such that
\[
\bu(\bx) = \left[|x_1|, |x_2|, \ldots, |x_N|\right]^T.
\]
\noindent
In this case, the loss function can be written as
\begin{align}
\mathcal{L}(\bz, \bg) &= (\bu(\bz) - \bu(\bg))^T \bW (\bu(\bz) - \bu(\bg)).\label{eq:nwls1}
\end{align}
\noindent Now, the gradient of loss with respect to the target model is then
\[
\nabla_\theta L = 2\left(\bU_r\bG_r + \bU_i\bG_i\right)^T \bW(\bu(\bg) - \bu(\bz)),
\]
\noindent where
\begin{align*}
\bU_r &= \frac{\partial \bu(\bg)}{\partial \bg_r},\nonumber\\
\bU_i &= \frac{\partial \bu(\bg)}{\partial \bg_i}.\nonumber
\end{align*}
\noindent Note that $\bU_r$ is a diagonal matrix with the $n^{\textrm{th}}$ entry down the diagonal being
\[
\left[\bU_r\right]_{nn} = \frac{\Re\left\{g(b_n, \bell)\right\}}{\left|g(b_n, \bell)\right|},
\]
\noindent and, likewise, the $\bU_i$ is a diagonal matrix with the $n^{\textrm{th}}$ entry down the diagonal being
\[
\left[\bU_i\right]_{nn} = \frac{\Im\left\{g(b_n, \bell)\right\}}{\left|g(b_n, \bell)\right|}.
\]

\subsection{Estimation Bounds}
Because the point scattering model is assumed to be differentiable (i.e., $\bG$ exists), the Cram\'{e}r-Rao lower bound (CRLB) can be directly computed if the range profile is observed in additive complex Gaussian noise \cite{KayEst93}. In this case, the CRLB for the model parameter vector is
\begin{equation}
E\left[\left(\btheta - \hat{\btheta}\right)\left(\btheta - \hat{\btheta}\right)^T\right] \succeq \bJ^{-1}(\bell),
\end{equation}
\noindent where $\bJ(\bell)$ is the Fisher information matrix defined as
\[
\bJ(\bell) = 2\Re\left\{\bG^H(\bell)\bR^{-1}\bG(\bell)\right\}.
\]

If the noise is independent across range profiles, the CRLB for multiple range profiles is simply
\begin{equation}
E\left[\left(\btheta - \hat{\btheta}\right)\left(\btheta - \hat{\btheta}\right)^T\right] \succeq \left(\displaystyle\sum_{m = 1}^{M}\bJ(\bell_m)\right)^{-1}.
\end{equation}

\section{Gradients}\label{sec:gradients}

The goal of this section is to derive the form of the Jacobian of a range profile, $\bG$, generated by a point scattering model. As discussed above (8), a range profile resulting from a point scattering model can be written as
\begin{align}
g(r, \bell) &= \displaystyle\sum_{n = 1}^N \gamma_n a_n R_{xx}(2r/c + 2\bp_n^T\bell/c).
\end{align}
\noindent By the definition of the total parameter vector, $\btheta$, the partial derivative of the range profile is
\[
\frac{\partial g(r, \bell)}{\partial \btheta} = \left[\frac{\partial g(r, \bell)}{\partial \btheta_1},
\frac{\partial g(r, \bell)}{\partial \btheta_2},
\ldots,
\frac{\partial g(r, \bell)}{\partial \btheta_N}\right].
\]
\noindent As each scatterer's position and amplitude only depend on their own parameter vector, $\btheta_n$, it can be observed that
\[
\frac{\partial g(r, \bell)}{\partial \btheta_n} = \frac{\partial}{\partial \btheta_n}\gamma_n a_n R_{xx}(2r/c + 2\bp_n^T\bell/c).
\]

Now, using the product rule for differentiation, it is possible to rewrite the partial derivative of the range profile as
\begin{align}
\frac{\partial g(r, \bell)}{\partial \btheta_n} &= \frac{\partial\gamma_n}{\partial \btheta_n}a_n R_{xx}(2r/c + 2\bp_n^T\bell/c)~+\nonumber\\
&\phantom{=}~\gamma_n \frac{\partial a_n}{\partial \btheta_n} R_{xx}(2r/c + 2\bp_n^T\bell/c)~+\nonumber\\
&\phantom{=}~\gamma_n a_n \frac{\partial R_{xx}(2r/c + 2\bp_n^T\bell/c)}{\partial \btheta_n}.\label{eq:prof_grad}
\end{align}
\noindent Therefore, to calculate the gradient of the range profile, $g(r, \bell)$, one needs to derive the gradients with respect to the phase delay, $\gamma_n$, complex weights, $a_n$, and autocorrelation function $R_{xx}(\tau)$.

Using the definition of the phase delay, $\gamma_n$, from (\ref{eq:gamma}), the partial derivatives with respect to the model parameter vector are
\begin{align}
\frac{\partial\gamma_n}{\partial \btheta_n} &= \frac{j4\pi f_c \gamma_n}{c}\bP^T_n\bell,\nonumber
\end{align}
\noindent where $\bP_n$ is the Jacobian of the $n^{\textrm{th}}$ point scatterer's position with respect to the parameter vector, $\btheta_n$, such that
\[
\bP_n = \frac{\partial \bp_n}{\partial \btheta_n}.
\]

As the autocorrelation term depends on the position of the point scatterer, the partial derivative of it can be expanded as
\begin{align}
\frac{\partial R_{xx}(2r/c + 2\bp_n^T\bell/c) }{\partial \btheta_n} &= \frac{2}{c}\frac{d R_{xx}(\tau)}{d \tau}\bP^T_n\bell.
\end{align}

Therefore, the components needed for calculating the partial derivatives of the range profile are:
\begin{itemize}
\item Scatterer amplitude derivatives with respect to parameter vector $\btheta_n$:
\[
\ba_n = \frac{\partial a_n}{\partial \btheta_n}
\]
\item Jacobians of scatterer position with respect to parameter vector $\btheta_n$:
\[
\bP_n = \frac{\partial \bp_n}{\partial \btheta_n}
\]
\item Derivative of waveform autocorrelation with respect to lag:
\[
\frac{d R_{xx}(\tau)}{d \tau}
\]
\end{itemize}
\noindent To exhibit the construction and utilization of a differentiable point scattering model, several fundamental scatterers and a common radar waveform will now be introduced along with their contributions to the data gradient Jacobian, \(\bG\).

\subsection{Fixed Amplitude Scatterer}
A scatterer with fixed amplitude is constant over all viewing geometries and only defined by a complex scattering coefficient $S = S_r + j S_i$. In this case, the parameter vector includes the real and imaginary parts of the scattering coefficient, i.e., $\btheta_{n} = [S_r, S_i]$. Thus, the amplitude is defined by
\begin{align}
a_n &= S.\nonumber
\end{align}
\noindent Further, the gradient entries with respect to $\btheta_{n}$ are
\begin{align}
\frac{\partial a_n}{\partial S_r } &= 1,\nonumber\\
\frac{\partial a_n}{\partial S_i } &= j.\nonumber
\end{align}
\noindent Compactly, this can be written as $\ba_n = [1, j]$.

\subsection{Fixed Position Scatterer}
A scatterer defined with a fixed position in cylindrical coordinates has a location with no dependence on the geometry vector $\bell$ and a parameter vector of $\btheta_{n} = [r_s, \phi_s, z_s]^T$ where $r_s$ is the radial component, $\phi_s$ is the azimuthal component, $z_s$ is the same as the $z$ component in a Cartesian system. More specifically, the relationship with the Cartesian point $\bp_n$ can be summarized as:
\begin{align}
p_x &= r_s\cos(\phi_s),\nonumber\\
p_y &= r_s\sin(\phi_s),\nonumber\\
p_z &= z_s,\nonumber
\end{align}
\noindent and
\begin{align}
r_s &= \sqrt{p_x^2 + p_y^2},\nonumber\\
\phi_s &= \tan^{-1}\left(p_y/p_x\right),\nonumber\\
z_s &= p_z.\nonumber
\end{align}
\noindent Then, positional Jacobian, $\bP_n$, is
\[
\bP_n = \left[\begin{array}{ccc}
\cos(\phi_s)&-r_s\sin(\phi_s)&0\\
\sin(\phi_s)&r_s\cos(\phi_s)&0\\
0&0&1
\end{array}\right].
\]

\subsection{Slipping Scatterer}
A slight variation of a fixed position scatterer is a \emph{slipping scatterer}, whose position always lies on the closest point on a ring whose normal is aligned with the $z$-axis. In this case, the parameter vector is $\btheta_{n} = [r_s, z_s]^T$ where $r_s$ is the radial component and $z_s$ is the same as the $z$ component in a Cartesian system. The entries of the position Jacobian, $\bP_n$, are the same as the fixed position scatterer except the azimuthal component, $\phi_s$, is now dictated by the line-of-sight vector, $\bell$, such that
\begin{align}
\phi_s &= \tan^{-1}\left(\ell_y/\ell_x\right).\nonumber
\end{align}
\noindent Then, positional Jacobian, $\bP_n$, is
\[
\bP_n = \left[\begin{array}{cc}
\cos(\phi_s)&0\\
\sin(\phi_s)&0\\
0&1
\end{array}\right].
\]

\subsection{Spherical Scatterer}
A spherical scatterer defined with a fixed position in cylindrical coordinates has a location with no dependence on the geometry vector $\bell$ and a parameter vector of $\btheta_{n} = \rho_s$ where $\rho_s$ is the radius of the sphere. More specifically, the relationship with a Cartesian point can be summarized as:
\begin{align}
p_x &= -\rho_s\ell_x,\nonumber\\
p_y &= -\rho_s\ell_y,\nonumber\\
p_z &= -\rho_s\ell_z,\nonumber
\end{align}
\noindent and
\begin{align}
\rho_s &= \sqrt{p_x^2 + p_y^2 + p_z^2}.\nonumber
\end{align}
\noindent Then, the positional Jacobian, $\bP_n$, is
\[
\bP_n = \left[\begin{array}{c}
-\ell_x\\
-\ell_y\\
-\ell_z
\end{array}\right].
\]

\begin{subsection}{Linear Frequency Modulated Waveform}

One of the most common waveforms used for radar measurements is the linear frequency modulated (LFM) signal \cite{Levanon88}. It is characterized by having a linear time-frequency relationship such that the instantaneous frequency, $f(t)$, is
\begin{equation}
f(t) = f_0 + \alpha t,\label{eq:lfm_tf}
\end{equation}
\noindent where $\alpha$ is the frequency rate of change. To get the analytic form of the waveform, the instantaneous phase, $\phi(t)$, can be obtained by integrating (\ref{eq:lfm_tf}) to get
\begin{align}
\phi(t) &= 2\pi\int_{0}^t f(\tau)d\tau,\nonumber\\
&= \pi\alpha t^2 + 2\pi f_0 t + \phi_0,
\end{align}
\noindent for some initial phase offset $\phi_0$. Thus, the waveform can be written as
\begin{align}
x(t) &= A e^{j \phi(t)},\nonumber\\
&= A e^{j \left(\pi\alpha t^2 + 2\pi f_0 t + \phi_0\right)},
\end{align}
\noindent where $A$ is the waveform amplitude, and the signal is defined over a time interval $t \in \left[0, T\right)$ for a duration $T$. The autocorrelation function as a function of time lag, $R_{xx}(\tau)$, can be derived as
\begin{align}
R_{xx}(\tau) &= \int_{-\infty}^{\infty}x(t)x^*(t - \tau)dt,\nonumber\\
&= A^2\int_{0}^{T}e^{j\phi(t) - j\phi(t - \tau)}dt,\nonumber\\
&= A^2e^{j2\pi f_0 \tau - j\pi\alpha\tau^2}\int_{0}^{T}e^{j2\pi\alpha t \tau}dt.\label{eq:lfm_auto1}
\end{align}
\noindent The integral portion of (\ref{eq:lfm_auto1}) can be simplified as
\begin{align}
\int_{0}^{T}e^{j2\pi\alpha t \tau}dt &= \frac{ e^{j2\pi T \alpha \tau} - 1}{j2\pi\alpha \tau},\nonumber\\
&= e^{j\pi T \alpha \tau}\left(\frac{e^{j\pi T \alpha \tau} - e^{-j\pi T \alpha \tau}}{j2\pi \alpha \tau}\right),\nonumber\\
&= e^{j\pi T \alpha \tau}\left(\frac{\sin\left(\pi T \alpha \tau\right)}{\pi \alpha \tau}\right).\nonumber
\end{align}
\noindent Thus, the complete autocorrelation function can be written as
\begin{equation}
R_{xx}(\tau) = A^2\frac{\sin\left(\pi T \alpha \tau\right)}{\pi \alpha \tau}e^{j\nu(\tau)},\label{eq:lfm_auto2}
\end{equation}
\noindent where $\nu(\tau)$ is
\[
\nu(\tau) = \pi\left(2 f_0 + T\alpha\right)\tau - \pi\alpha\tau^2.
\]
\noindent Thus, using the product rule, the partial derivative with respect to $\tau$ is
\begin{align}
\frac{d}{d \tau}R_{xx}(\tau)&= A^2 \frac{d}{d \tau}\frac{\sin\left(\pi T \alpha \tau\right)}{\pi \alpha \tau} e^{j\nu(\tau)}~+\nonumber\\
&\phantom{= }~j A^2 \frac{\sin\left(\pi T \alpha \tau\right)}{\pi \alpha \tau} e^{j\nu(\tau)} \frac{d \nu(\tau)}{d \tau},\nonumber
\end{align}
\noindent where
\begin{align}
\frac{d}{d \tau}\frac{\sin\left(\pi T \alpha \tau\right)}{\pi \alpha \tau} &= \frac{T\cos\left(\pi T \alpha \tau\right)}{\tau} - \frac{\sin\left(\pi T \alpha \tau\right)}{\pi \alpha \tau^2},\nonumber
\end{align}
\noindent and
\[
\frac{d}{d \tau}\nu(\tau) = \pi\left(2 f_0 + T\alpha\right) - 2\pi\alpha \tau.
\]



\end{subsection}

\section{Results}\label{sec:results}

In this section, estimation using the proposed model will be studied. The true model used for this example consists of three point scatterers:
\begin{itemize}
    \item Fixed amplitude of $S = 1$; fixed position of $(r_s, \phi_s, z_s) = (0.5, 0, 2)$,
    \item Fixed amplitude of $S = 2$; fixed position of $(r_s, \phi_s, z_s) = (0, \pi/8, -2)$,
    \item Fixed amplitude of $S = 0.5$; slipping position of $(r_s, z_s) = (0, 0.1)$.
\end{itemize}

Model optimization will be performed using gradient descent \cite{Chong13} and a line search performed at each iteration to determine the optimal step size. More specifically, the estimate of the model parameter vector at the $k^{\textrm{th}}$ iteration, $\hat{\btheta}^{(k)}$ is given by
\[
\hat{\btheta}^{(k)} = \hat{\btheta}^{(k - 1)} - \eta\nabla_{\theta} L,
\]
\noindent where $\eta$ is defined as
\[
\eta = \argmin_{\eta'} L\left(\bz, \bg\left(\hat{\btheta}^{(k - 1)} - \eta' \nabla_{\theta} L\right)\right).
\]
\noindent The initial guess, $\hat{\btheta}^{(0)}$, is assumed to be given as the following:
\begin{itemize}
    \item Fixed amplitude of $S = 1.01$; fixed position of $(r_s, \phi_s, z_s) = (0.6, 0, 2.1)$,
    \item Fixed amplitude of $S = 1.9$; fixed position of $(r_s, \phi_s, z_s) = (0, \pi/8 + 0.01, -2.1)$,
    \item Fixed amplitude of $S = 0.51$; slipping position of $(r_s, z_s) = (0.1, 0.1)$.
\end{itemize}
\noindent There will be three types of estimation considered: noncoherent, coherent, and sequential. Noncoherent estimation uses noncoherent weighted squared error, coherent estimation uses coherent weighted squared error, and sequential estimation uses noncoherent estimation to get a coarse model estimate and then uses coherent estimation to refine the answer.

\begin{figure}
  \centering
  \includegraphics[width=0.95\columnwidth]{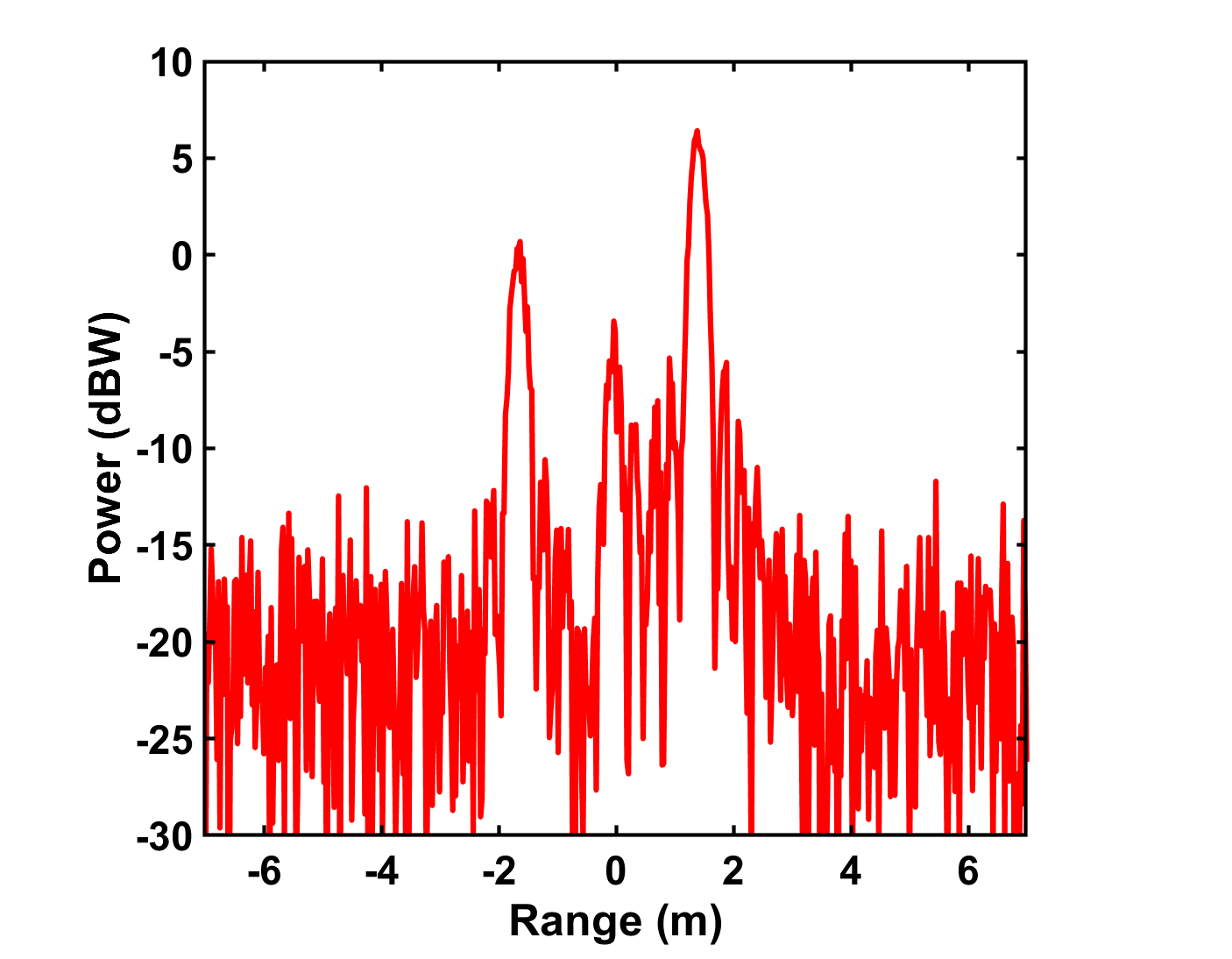}
  \caption{Range profile of target using LFM waveform with 500 MHz bandwidth and center frequency $f_c = 3~\mathrm{GHz}$.}\label{fig:res1_prof}
\end{figure}

\begin{figure}
  \centering
  \includegraphics[width=0.95\columnwidth]{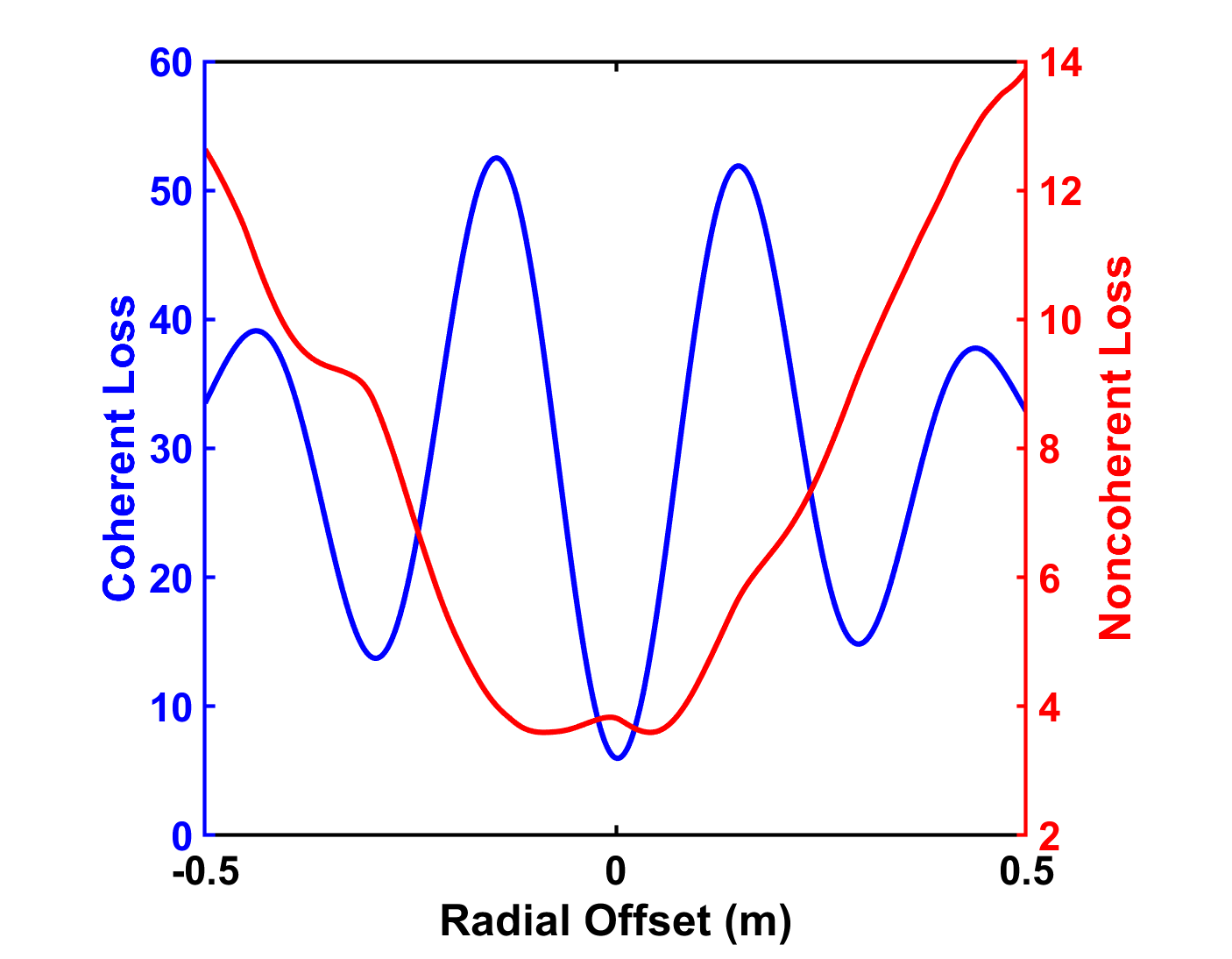}
  \caption{Coherent and noncoherent loss versus fixed scatterer radial position offset.}\label{fig:res1_loss}
\end{figure}

\begin{figure}
  \centering
  \includegraphics[width=0.95\columnwidth]{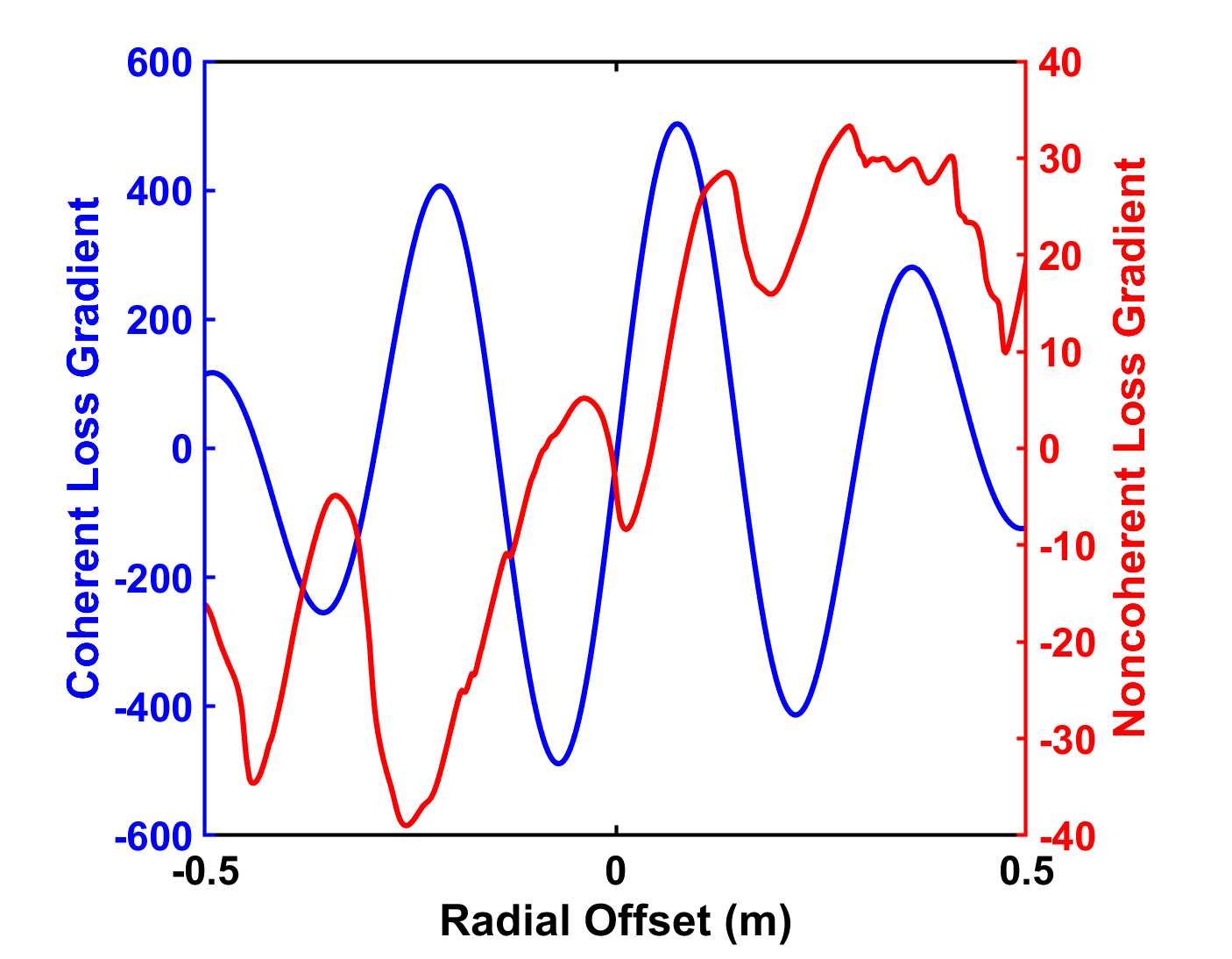}
  \caption{Gradients of coherent and noncoherent loss with respect to fixed scatterer radial position.}\label{fig:res1_dloss}
\end{figure}

In the first example, a point scattering model will be fit to a single range profile using different loss functions. In Figure \ref{fig:res1_prof}, the observed noise range profile is given; the noise variance is given as $\sigma^2 = 0.01$. Before fitting a model, the loss function behavior will be investigated. To this end, the radial position, $r_s$, of the first scatterer will be varied from the true value (while keeping all other parameters at their true values) and the loss function calculated. In Figure \ref{fig:res1_loss}, the coherent and noncoherent weighted squared error is shown as a function of the offset of the radial position, $r_s$, of the first scatterer. It can be seen that the coherent loss is much more sensitive to changes in parameter value; this is due to the change in phase delay caused by shifting the scatterer in range. This effect also leads to the sinusoidal behavior of the coherent loss function. In Figure \ref{fig:res1_dloss}, the gradients of the loss functions with respect to the radial position, $r_s$, of the first scatterer are shown. Like the loss functions in Figure \ref{fig:res1_loss}, the gradients show that the coherent loss requires a more accurate initial guess to successfully perform gradient descent, but it also provides a much higher accuracy of estimate. This intuition provides the motivation of using sequential estimation: noncoherent estimation refines a rough initial guess to a coarse model estimation, then coherent estimation is used to refine the estimate.

\begin{figure}
  \centering
  \includegraphics[width=0.95\columnwidth]{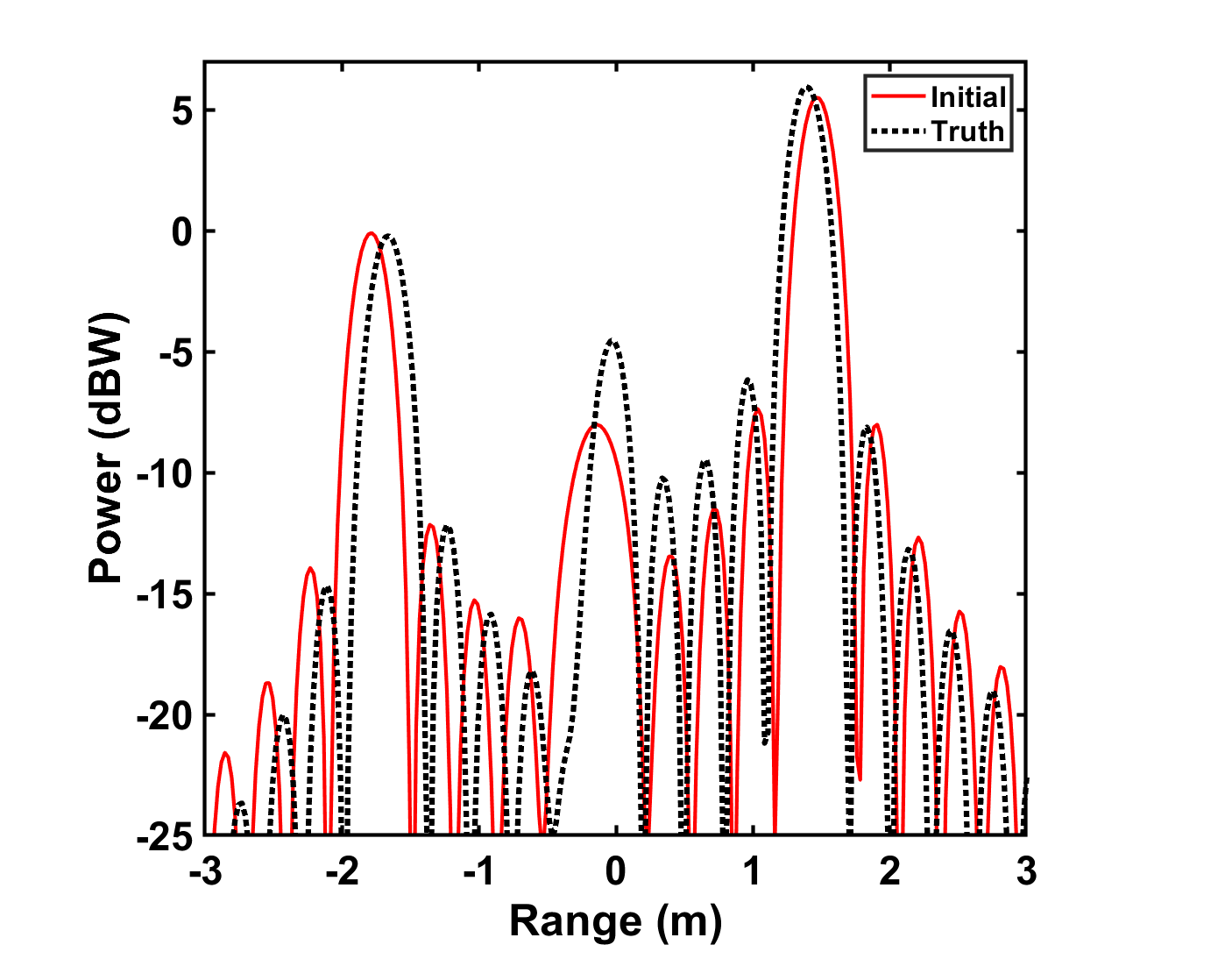}
  \caption{True range profile and range profile resulting from initial guess.}\label{fig:res1_initial}
\end{figure}

\begin{figure}
  \centering
  \includegraphics[width=0.95\columnwidth]{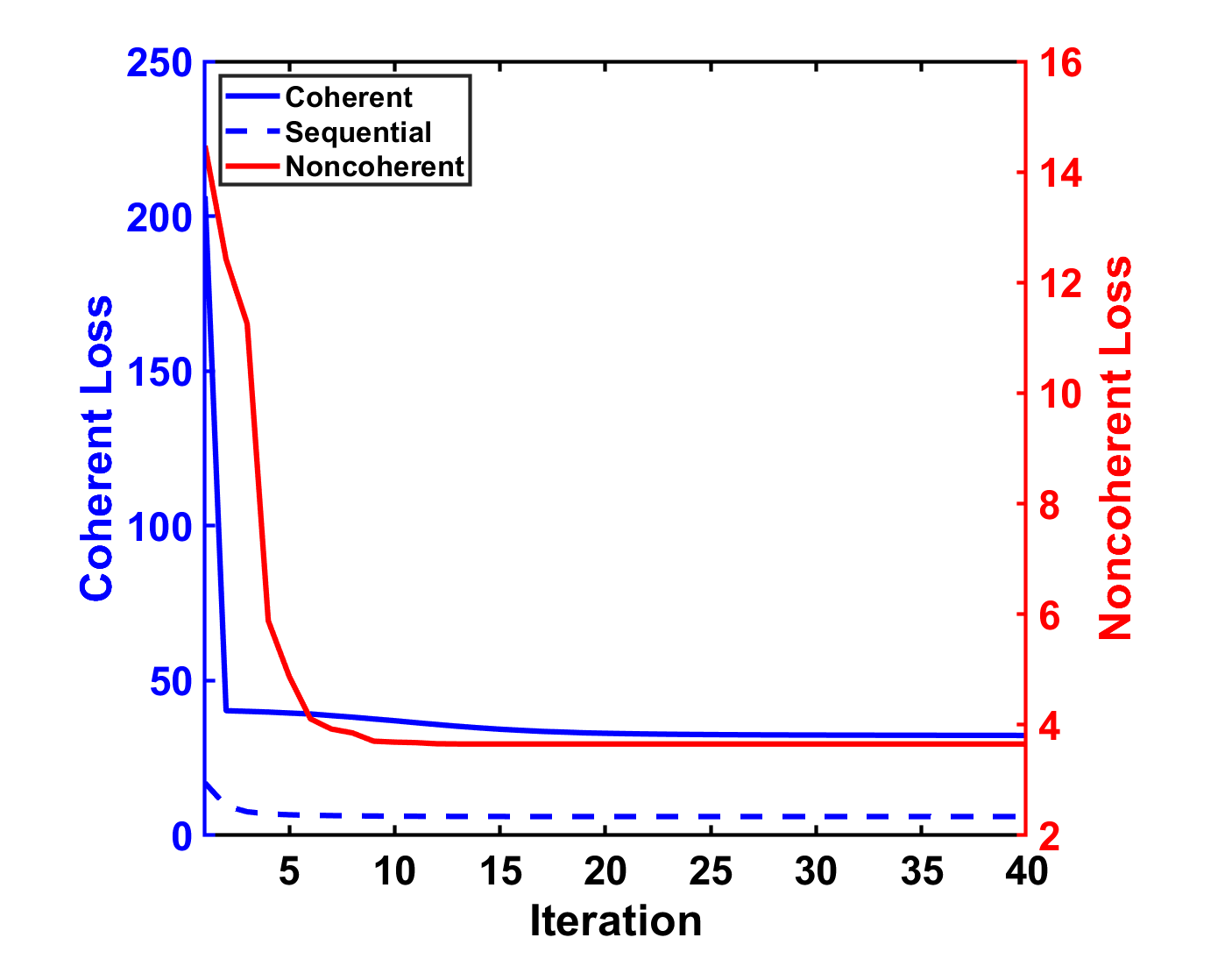}
  \caption{Loss value evolution over iterations of gradient descent.}\label{fig:res1_iter}
\end{figure}

\begin{figure}
  \centering
  \includegraphics[width=0.95\columnwidth]{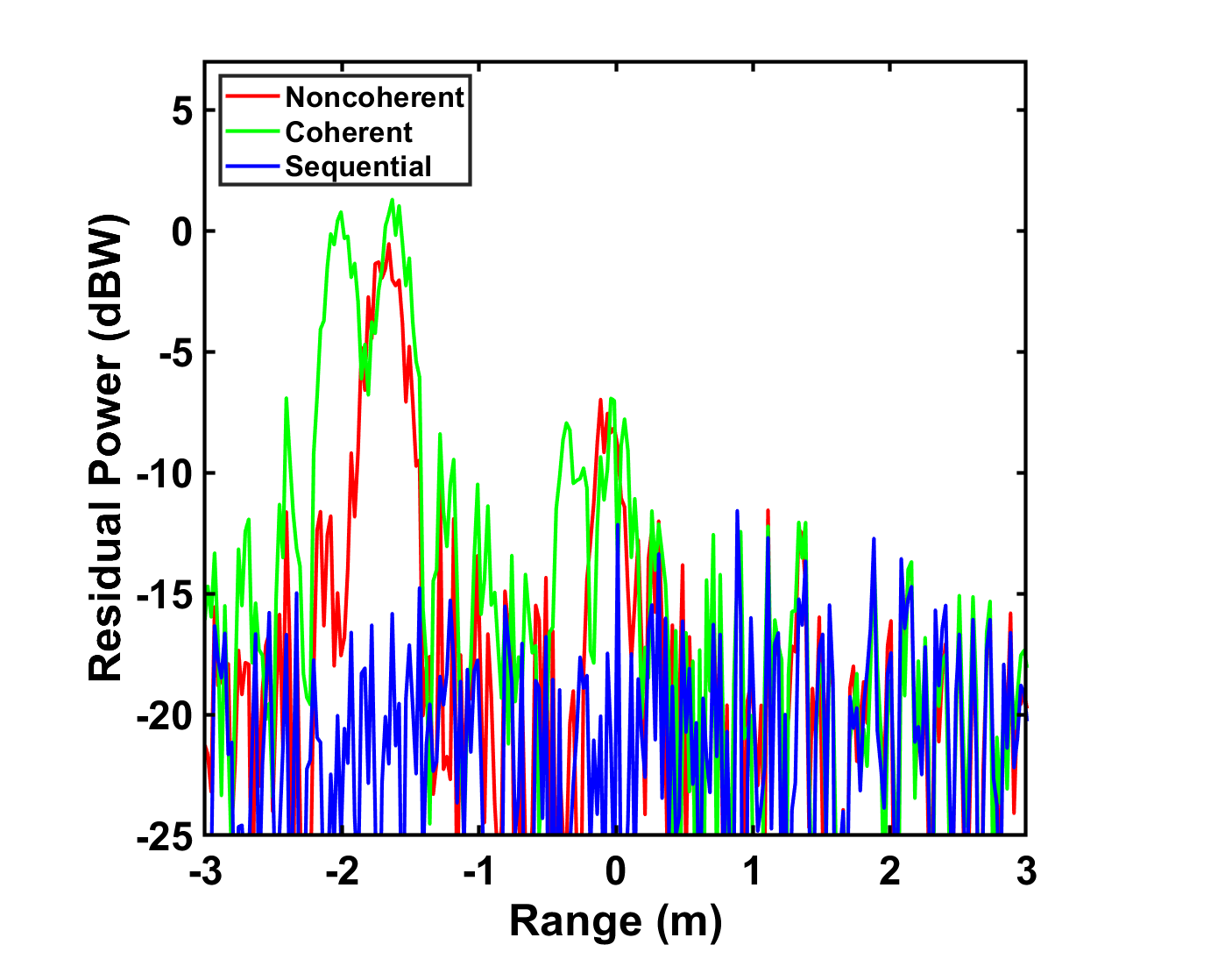}
  \caption{Power of range profile residuals after model optimizations.}\label{fig:res1_res}
\end{figure}

Now, focus will be shifted to optimization of the model to fit the range profile given in Figure \ref{fig:res1_prof}. The range profile produced by the initial guess at model parameters is compared to the true range profile in Figure \ref{fig:res1_initial}. The evolution of the loss functions versus iteration of gradient descent are shown in Figure \ref{fig:res1_iter}. It can be seen that coherent estimation converged to a local minimum that does not correspond to the global optimal; however, sequential estimation utilized the first pass of noncoherent estimation to avoid this phenomenon. The instantaneous power of the range profile residuals are shown in Figure \ref{fig:res1_res}. It can be seen that the residual of sequential estimation is on the order of the noise power (-20 dBW); thus, the model has been fit effectively.

\begin{figure}
  \centering
  \includegraphics[width=0.99\columnwidth]{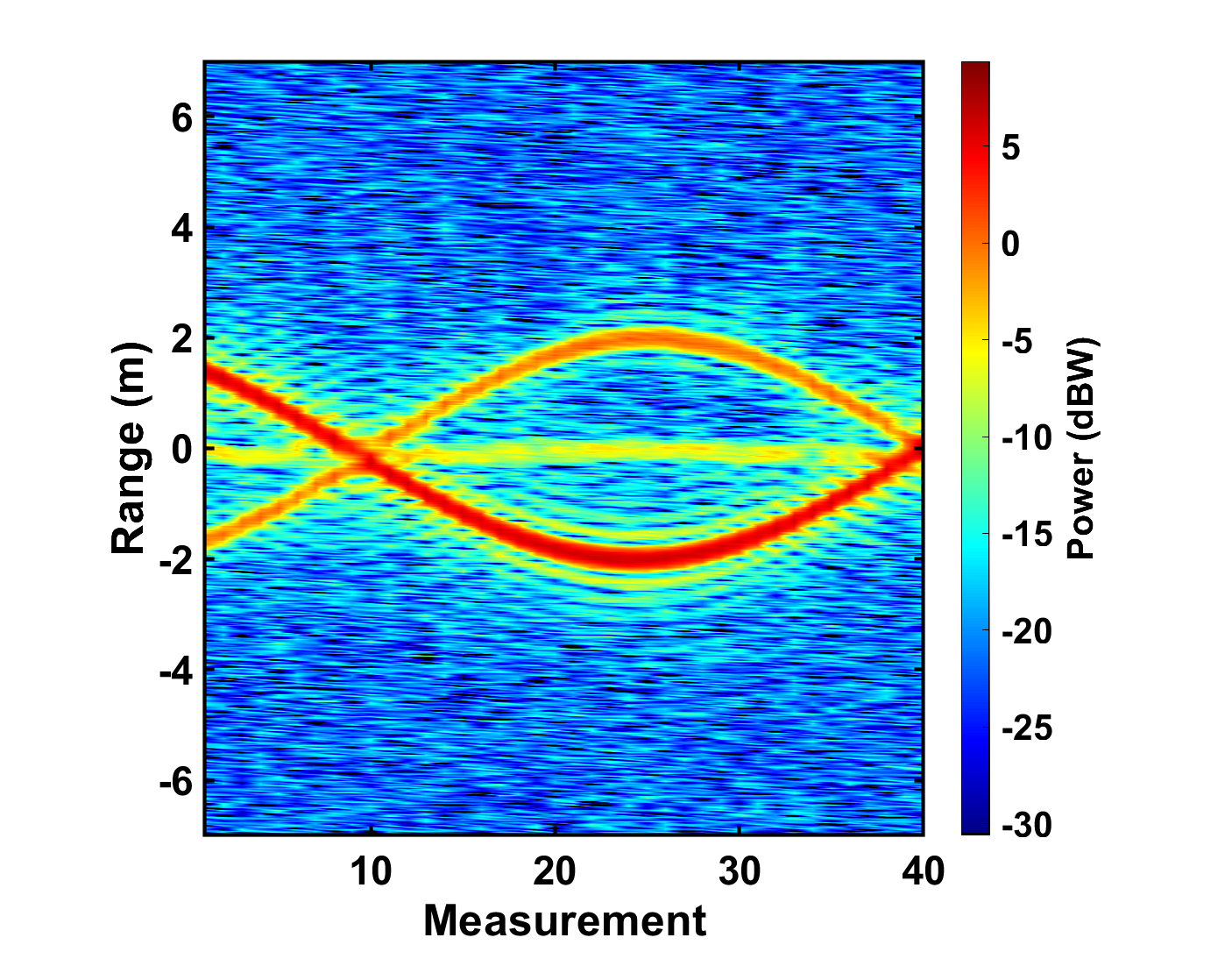}
  \caption{Static pattern of model using LFM waveform with 500 MHz bandwidth and center frequency $f_c = 3~\mathrm{GHz}$; noise variance is $\sigma^2 = 0.01$.}\label{fig:res2_stat}
\end{figure}

To extend past a single range profile, optimizing over a set of range profiles is now considered. More specifically, fitting a model to a \emph{static pattern} will now be studied. A static pattern is a collection of range profiles at different viewing geometries; the adjective ``static'' is used to emphasize that the target is very slowly moving in each observation, e.g., the Doppler effect is negligible. In Figure \ref{fig:res2_stat}, a noise static pattern is given using the same observation parameters as in the previous result; the main difference is that the viewing angle now varies for each measurement.

\begin{figure}
  \centering
  \includegraphics[width=0.99\columnwidth]{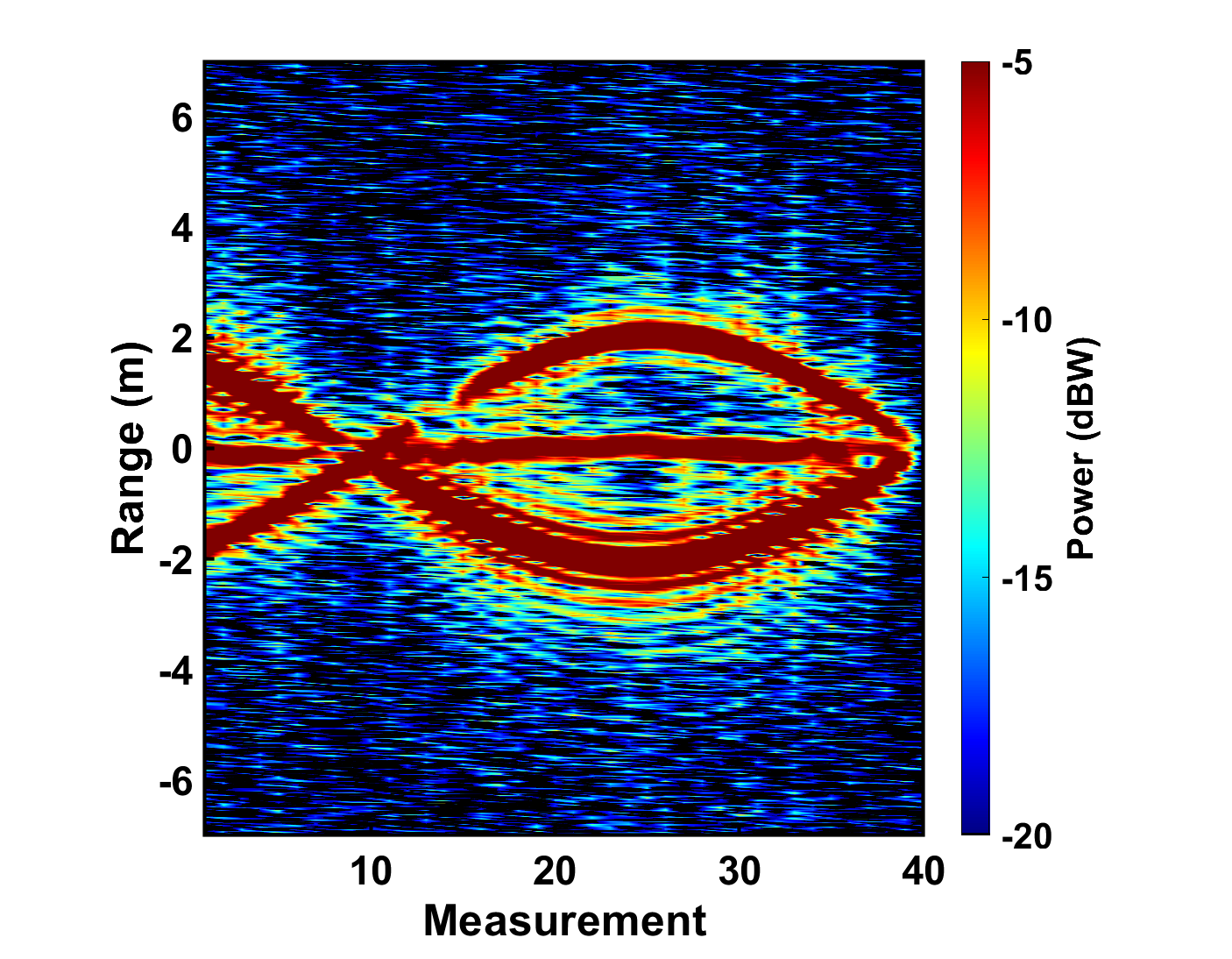}
  \caption{Power of residual image using initial guess.}\label{fig:res2_initial}
\end{figure}

In Figure \ref{fig:res2_initial}, the instantaneous power of the residual image due to the initial guess at model parameters is given. It can be noted that there is significant difference between the hypothesized image and the noisy observation.

\begin{figure}
  \centering
  \includegraphics[width=0.95\columnwidth]{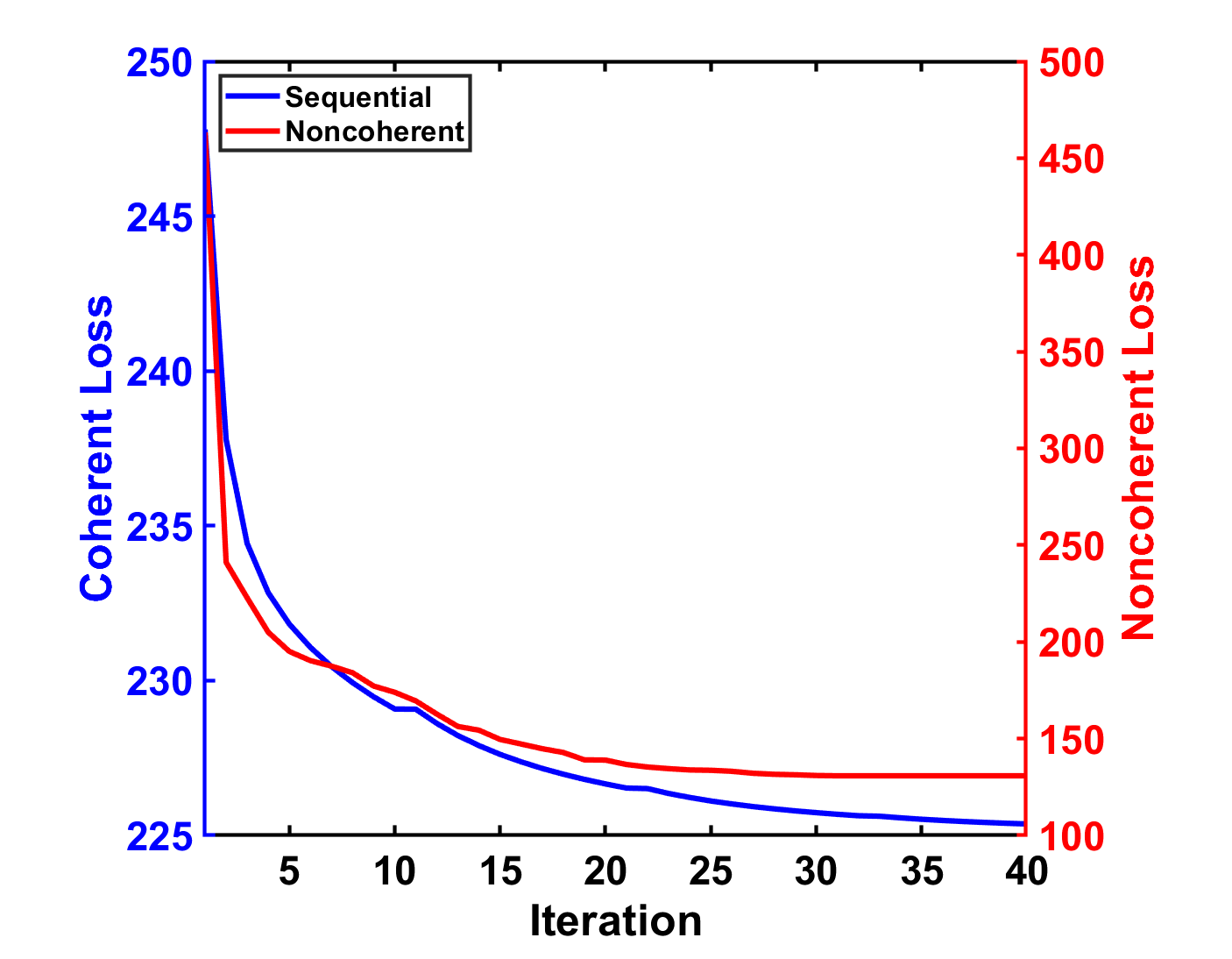}
  \caption{Loss value evolution over iterations of gradient descent.}\label{fig:res2_iter}
\end{figure}

\begin{figure}
  \centering
  \includegraphics[width=0.99\columnwidth]{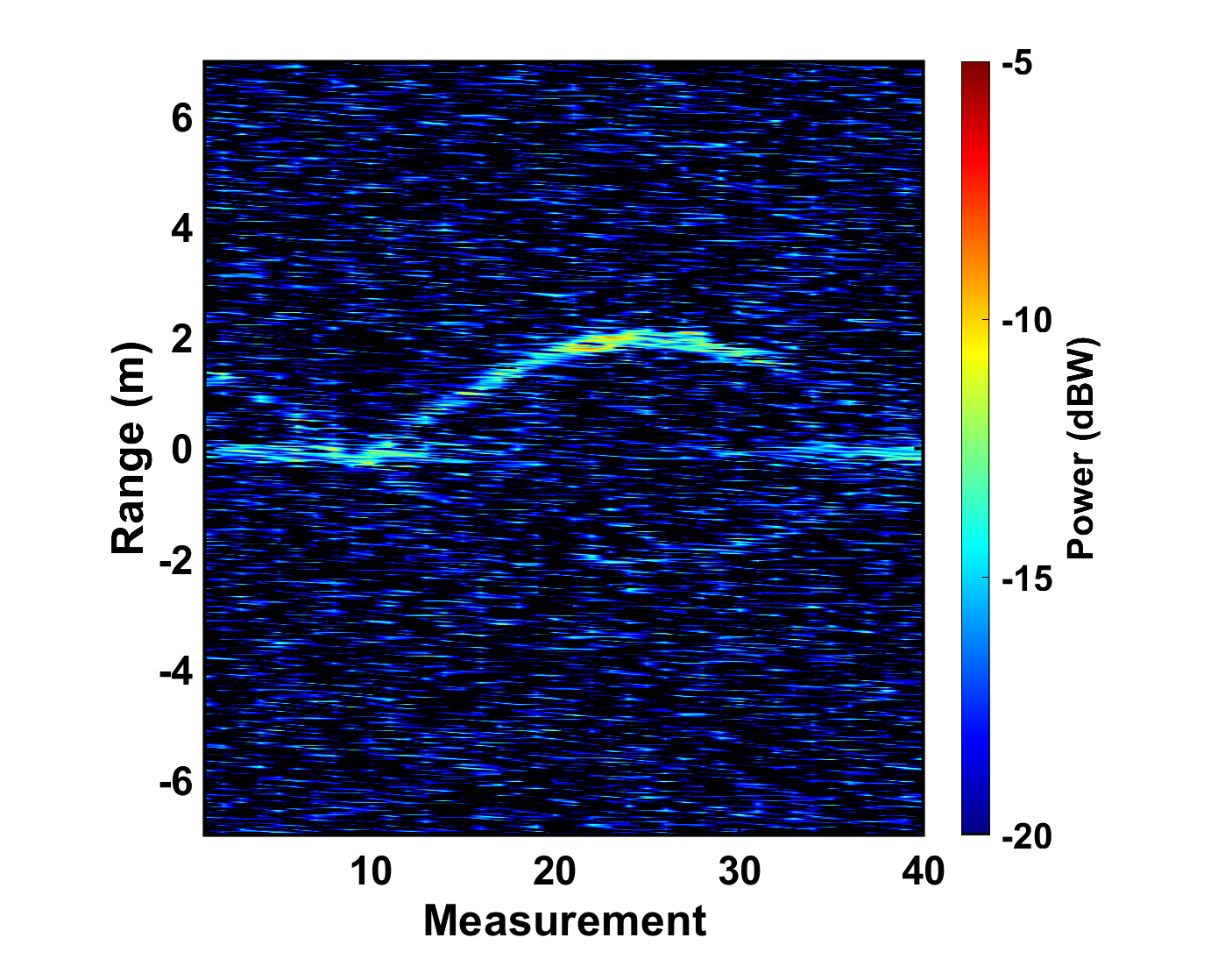}
  \caption{Power of residual image after noncoherent model optimization.}\label{fig:res2_non}
\end{figure}

\begin{figure}
  \centering
  \includegraphics[width=0.99\columnwidth]{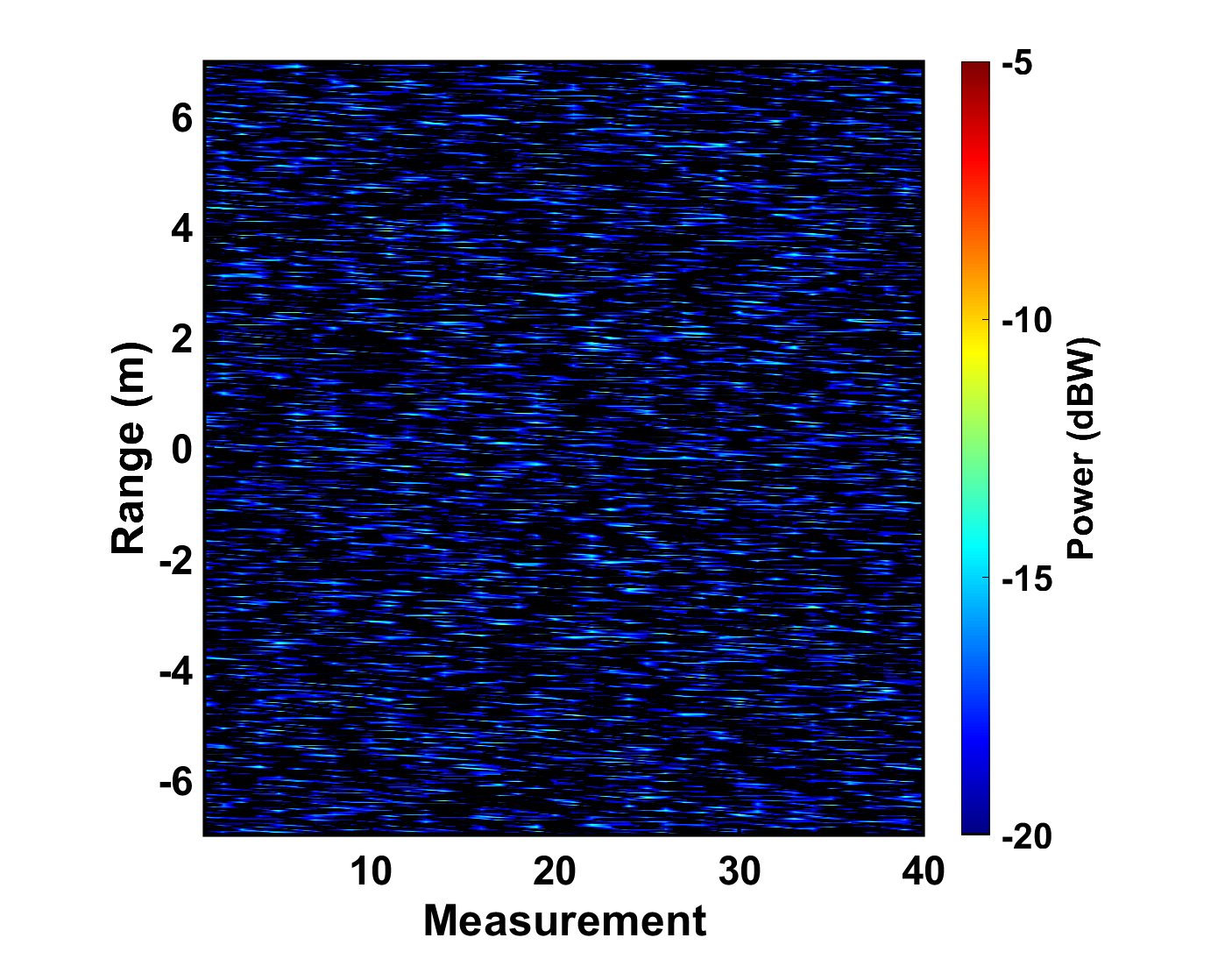}
  \caption{Power of residual image after sequential model optimization.}\label{fig:res2_coh}
\end{figure}

In Figure \ref{fig:res2_iter}, the loss function values at each iterations are shown for noncoherent and sequential estimation. The instantaneous powers for the residual images are shown in Figures \ref{fig:res2_non} and \ref{fig:res2_coh}. It can be seen that, as with the single range profile, noncoherent estimation provides a coarse model estimate that is then successfully refined by coherent estimation (shown in the sequential estimate residual).

\section{Conclusion}

This work proposes a set of fundamental point scattering components that can be used in combination to form differentiable scattering models. The models, in turn, can then employ gradient-based optimization methods to enable fast optimization for fitting observed data. This offers efficient radar target model manipulation, which is an important step in the process of target characterization.

\section{Distribution}
DISTRIBUTION STATEMENT A. Approved for public release. Distribution is unlimited.

This material is based upon work supported by the Under Secretary of Defense for Research and Engineering under Air Force Contract No. FA8702-15-D-0001. Any opinions, findings, conclusions or recommendations expressed in this material are those of the author(s) and do not necessarily reflect the views of the Under Secretary of Defense for Research and Engineering.

\FloatBarrier

\bibliographystyle{IEEEtran}
\bibliography{main}

\end{document}

%% file: include.tex
\usepackage{amsmath}
\usepackage{amssymb}
\usepackage{amsthm}
\usepackage{epsf}
\usepackage{epsfig}
\usepackage{multirow}
\usepackage{scalefnt}
\usepackage{subcaption}
\usepackage{siunitx}
\usepackage[normalem]{ulem}
\usepackage{xcolor}
\usepackage{enumerate}
\usepackage{empheq} 

%% file: custom_commands.tex
\newtheorem{definition}{Definition}

  \def\cC{{\mathcal{C}}}

 \def\cN{{\mathcal{N}}}

\def\argmin{\mathop{\mathrm{argmin}}}

\def\Re{\mathop{\mathrm{Re}}}
\def\Im{\mathop{\mathrm{Im}}}

   \def\btheta{{\pmb{\theta}}}

\def\bzeros{{\pmb{0}}}

\def\ba{{\mathbf{a}}}   
  \def\bg{{\mathbf{g}}} 
   \def\bell{{\pmb{\ell}}}
   \def\bp{{\mathbf{p}}}
   
\def\bu{{\mathbf{u}}}  \def\bw{{\mathbf{w}}} \def\bx{{\mathbf{x}}}
 \def\bz{{\mathbf{z}}}  

  \def\bG{{\mathbf{G}}} 
 \def\bJ{{\mathbf{J}}}  
   \def\bP{{\mathbf{P}}}
 \def\bR{{\mathbf{R}}}  
\def\bU{{\mathbf{U}}}  \def\bW{{\mathbf{W}}}
